\documentclass[twocolumn,showpacs,preprintnumbers,amsmath,amssymb]{revtex4}

\usepackage{graphicx}  
\usepackage{dcolumn}
\usepackage{bm}

\begin{document}
\title{Crossover behavior and multi-step relaxation in a schematic model of the cut-off glass transition}
\date\today
\author{M.~J.~Greenall}
\affiliation{SUPA, School of Physics, University of Edinburgh,
JCMB King's Buildings, Edinburgh EH9 3JZ, U.K.}
\affiliation{School of Physics and Astronomy, University of Leeds, Leeds LS2 9JT, U.K.}
\author{M.~E.~Cates}
\affiliation{SUPA, School of Physics, University of Edinburgh,
JCMB King's Buildings, Edinburgh EH9 3JZ, U.K.}

\begin{abstract}
We study a schematic mode-coupling model in which the ideal glass transition is cut off by a decay of the quadratic coupling constant in the memory function. (Such a decay, on a time scale $\tau_I$, has been suggested as the likely consequence of activated processes.) If this decay is complete, so that only a linear coupling remains at late times, then the $\alpha$ relaxation shows a temporal crossover from a relaxation typical of the unmodified schematic model to a final strongly slower-than-exponential relaxation. This crossover, which differs somewhat in form from previous schematic models of the cut-off glass transition, resembles light-scattering experiments on colloidal systems, and can exhibit a `slower-than $\alpha$' relaxation feature hinted at there. We also consider what happens when a similar but incomplete decay occurs, so that a significant level of quadratic coupling remains for $t\gg\tau_I$. In this case the correlator acquires a third, weaker relaxation mode at intermediate times. This empirically resembles the $\beta$ process seen in many molecular glass formers. It disappears when the initial as well as the final quadratic coupling lies on the liquid side of the glass transition, but remains present even when the final coupling is only just inside the liquid (so that the $\alpha$ relaxation time is finite, but too long to measure). Our results are suggestive of how, in a cut-off glass, the underlying `ideal' glass transition predicted by mode-coupling theory can remain detectable through qualitative features in dynamics.
\end{abstract}

\maketitle

\section{Introduction}

Many aspects of the slowing down of the dynamics of a liquid as it is cooled or compressed towards its glass transition are accurately captured by mode-coupling theory (MCT) \cite{leutheusser, bgs}. Although cast in terms of collective coordinates (Fourier components of density), MCT is often thought to model the formation of `cages' whereby a given particle is constrained by its neighbors in real space. However, the theory, at least in its simplest (`standard') form (SMCT), breaks down in the final approach to the glassy state (see, for example, Refs \cite{angell_rev, debenedetti_rev, reichman_rev}). In particular, MCT predicts a complete arrest of density fluctuations at long times \cite{leutheusser, bgs}; this {\em ideal glass transition} occurs at weaker coupling (higher temperatures and lower densities) than the experimental glass transition. The continued decay of fluctuations at stronger couplings is usually attributed to activated (`hopping') processes, not captured by SMCT \cite{goetze_rev}. In their simplest form, these could involve escape from a cage by activation over a local barrier, which would give an Arrhenius factor cutting off the divergence of the relaxation time. However, in many glasses, the relaxation time shows a stronger-than-Arrhenius dependence far into the regime beyond the ideal glass transition, suggesting a collective aspect to the hopping dynamics. Much current research addresses many-body activated processes at a supra-cage scale, often referred to as {\em dynamical heterogeneity} \cite{sollich_rev}.

Most recent efforts to understand these extra relaxation pathways fall into two broad categories: analysis of microscopic molecular dynamics simulations in the light of free-energy landscape ideas \cite{debenedetti_rev, stillinger_rev,sastry}, and study of coarse-grained models \cite{garrahan_chandler,xia_wolynes,geissler_reichman} aimed to directly address dynamical heterogeneity. Despite the new insights afforded by these methods, we believe that the success of MCT in describing the viscous liquid state means that one should not abandon attempts to extend its validity beyond the ideal glass transition. Indeed, several programmes along these lines have been carried out, either by avoiding some of the approximations of MCT \cite{mayer}, introducing new decay modes into the full MCT formalism (creating `extended MCT' or eMCT) \cite{goetze_emct}, or adding relaxation processes not easily described through mode-coupling to a simplified version of the theory (see e.g.\ Refs \cite{bagchi,schweizer_saltzman}).

In this paper we take a related route: we make a specific ad-hoc adjustment to a set of equations that, unmodified, are an accepted `schematic' representation of SMCT \cite{goetze_les_houches}. The adjustment we propose is inspired by recent theoretical work \cite{cates_ramaswamy} which addresses not just whether, but {\em how} activated processes
(`instantons') can violate the central approximation of SMCT. The latter involves factoring a four-point correlator into the product of two two-point correlators \cite{goetze_rev}. Cates and Ramaswamy argued that, in an instanton-dominated regime, the four-point correlator should be better approximated by a single two-point correlator. A crossover to a regime in which that approximation holds, in place of the MCT one, can gradually weaken the density-density coupling and hence can switch off the feedback that drives the arrest. This line of reasoning gives a cut-off glass transition, but one with a different mathematical structure from those proposed in connection with various forms of eMCT \cite{goetze_emct,das_mazenko,das_rev}.
The resulting model structure is also different from that found by multiplying either the MCT correlator or the MCT memory function by an independent decay function.

Below we explore in more depth the consequences of a crossover, on an `instanton' timescale $\tau_I$, from a quadratic to a linear dependence of four-point correlators on two-point correlators. This we do
within the framework of a schematic, single-mode model. The simplicity of this framework introduces several artificial features precluding any detailed comparison with experimental data; however, a wide range of glassy relaxation scenarios can be found within this simple model. Though not done here, this suggests that pursuit of similar ideas within a full (multi-mode, wavevector-dependent) MCT could be a fruitful avenue of future research.

Section \ref{mct} gives a sketch of the mode-coupling theory of the glass transition, and its schematic versions are outlined in Sec.\ \ref{schematic}. We introduce the phenomenology and existing theory of the relevant glassy relaxation scenarios in Sec.\ \ref{multistep} and the model with cut-off in Sec.\ \ref{cutoff}. Our numerical procedures are detailed in Sec.\ \ref{numerics}, and results (\ref{results}) and conclusions (\ref{conclusions}) follow.

\section{Mode-Coupling Theory of the Glass Transition}\label{mct}

We now present a brief outline of the mode-coupling theory of the glass transition, reviewed in depth elsewhere \cite{goetze_rev,goetze_exp,cummins_rev}.

As a glass-forming liquid is cooled or compressed, the viscosity and,
more generally, the relaxation times for density fluctuations
in the system increase rapidly. As the glass transition is reached, unequal-time density correlators  that would, in an ergodic material, decay to zero at late times are no longer able to do so on the timescale of the experiment. A proportion of the density fluctuations are then arrested or `frozen', although the system remains amorphous. 

The theory describes this through a feedback mechanism driven by couplings among density fluctuations, often thought to be associated with the {\em caging} of a particle by its neighbors. To quantify the effect, MCT uses the normalized autocorrelation functions of density fluctuations at wavevector $q$, defined as
\begin{equation}
\phi_q(t)=\langle\rho_q(t)\rho^*_q(0)\rangle/\langle|\rho_q|^2\rangle
\label{density_corr}
\end{equation}
where the angular brackets denote equilibrium averages. The $\phi_q$ evolve according to
\begin{equation}
\ddot{\phi}_q(t)+\gamma_q\dot{\phi}(t)+\Omega_q^2\phi_q(t)+\Omega_q^2\int^t_0m_q(t-t')\dot{\phi}_q(t')\,\mathrm{d}t'=0
\label{mcteom}
\end{equation}
This is derived from the Newtonian equations of motion for $N$ interacting particles  and represents their averaged dynamics exactly. 
(Alternatively, the second-derivative term may be dropped; the equation then describes a system of interacting Langevin particles with uncorrelated local noise terms. The latter can be used to describe a system of Brownian colloids in the absence of hydrodynamic interactions.)
The microscopic frequencies $\Omega_q$ may be calculated from the static properties of the liquid and $\gamma_q$ is a regular damping term separated from the memory kernel $m_q$ so that the latter describes only the dominant effects near the glass transition. Approximations must be made to allow $m_q$ to be calculated. The vital step in MCT is the identification of the density fluctuation pairs $\delta\rho_{q_1}\delta\rho_{q_2}$ (with $q_1+q_2=q$) as the major contributions to the memory kernel; those parts of $m_q$ uncorrelated with these are dropped. Further approximations, including notably the factorization of the four-point correlation functions arising from these product modes, lead to an expression for $m_q$ whose only input is the mean density $\rho$ and structure factor $S(q)=\langle\rho_q(0)\rho^*_q(0)\rangle$ of the equilibrium liquid state (see e.g.\ Ref.\ \cite{franosch}). As the static correlations described by $S(q)$ increase, the feedback becomes stronger, and the decay of $\phi_q(t)$ stretches out to longer times. Eventually, at the ideal glass transition $T_c(\rho)$ (or $\rho_c(T)$), the fluctuations never decay entirely, and $\lim_{t\to\infty}\phi_q(t)\equiv f_q >0$. (The residual value $f_q$ is the $q$-dependent nonergodicity parameter.) $S(q)$ remains regular at $T_c$: mode-coupling theory does not require a thermodynamic singularity to trigger the glass transition.

This approach has had remarkable success in modelling the detailed time- and wavevector-dependence of the $\phi_q(t)$ as the glass transition is approached, especially in colloidal fluids \cite{goetze_exp}, where it also provides a coherent description of a glass-to-glass transition between arrested phases dominated by inter-particle repulsion and attraction respectively \cite{dawson,bergenholtz,pham}. It successfully predicts the two-step decay of density correlations near the glass transition, where a fast initial relaxation towards a plateau (extending to infinite times at $T_c$) is followed by a slower-than-exponential final decay. Furthermore, several aspects of this relaxation are fixed by a single {\em exponent parameter} $\lambda$, calculated from $S(q)$.

Despite the success of mode-coupling theory in describing the viscous liquid state, agreement between the asymptotic decay laws of SMCT and experiment rapidly worsens at temperatures around $1.2T_g$ \cite{wiedersich_toluene,adichtchev_et_al,adichtchev_glycerol,blochowicz_et_al,roessler_beta,hansen,adichtchev_light,roessler_glycerol}. This problem is often attributed to the presence of additional relaxation processes not captured in the theory (sometimes referred to as {\em hopping} in molecular systems). This school of thought argues that SMCT captures the physics of an {\em ideal glass transition} (at $T_c\approx 1.2T_g$) which is avoided in reality but which would arise in a world where such additional processes did not exist; worsening agreement at $T\approx 1.2T_g$ is then taken to signify a qualitative change in transport mechanism from collective to hopping processes (see, for example \cite{sastry,roessler_beta,adichtchev_et_al,adichtchev_light,blochowicz_et_al,wiedersich_toluene}).

\section{Schematic MCT models of the glass transition}\label{schematic}
Many features of the full (wavevector-dependent) SMCT can be captured by simple schematic models. In these, eqn.\ \ref{mcteom} is reduced to
\begin{equation}
\partial^2_t\phi(t)+\nu\Omega^2\partial_t\phi(t)+\Omega^2\phi(t)+\Omega^2\int^t_0\mathrm{d}t\, m(t-\tau)\partial_\tau\phi(\tau)=0
\label{mcteom2}
\end{equation}
where all $q$-dependence has been dropped. This single correlator is usually associated with density fluctuations at the wavelength of the peak in the structure factor \cite{bgs}.

In contrast to the full theory, there is no clear microscopic prescription for the form of $m$ in schematic models. The simplest form for the memory kernel giving a reasonably realistic picture of the glass transition is $m(t)=v_2\phi(t)^2$, with $v_2$ an adjustable coupling parameter. Together with Eqn.\ \ref{mcteom2}, this defines the $\text{F}_2$, or Leutheusser, model \cite{leutheusser}. The quadratic term is a simple representation of the density-density coupling in the full SMCT. The $\text{F}_2$ model gives a discontinuous jump of the nonergodicity parameter $f\equiv\lim_{t\to\infty}\phi(t)$ at a critical value of the control parameter ($v_2=4$) (a {\em type B} transition), and also yields a good qualitative description of the characteristic two-step decay of $\phi(t)$ in the approach to the glass transition. Its ability to fit experimental data for $\phi(t)$ is limited by the fact that the detailed form of the two-step decay is rather firmly fixed, and in some respects is unrealistic. In particular, the height of the intermediate-time plateau $f^\text{c}$ in $\phi(t)$ is always $0.5$ in the liquid (although this, and hence also the long-time limit $f$, may be $>0.5$ in the glass), and the slow ($\alpha$) decay at long times is described by a simple exponential rather than by some slower function (most experimental data in this regime may be well fitted by stretched exponentials \cite{cummins_exp}).

The $\text{F}_2$ model can be made more flexible by the addition of a linear term to the memory kernel, giving $m(t)=v_1\phi(t)+v_2\phi(t)^2$. This model, called $\text{F}_{12}$ \cite{goetze_les_houches}, allows the plateau height $f^\text{c}$ and the form of the decay to and from the plateau to be adjusted, and also gives a (close to) stretched exponential form for the long-time relaxation. However, the functional form of the decay cannot be varied independently of the plateau height, which itself is restricted to $f^\text{c}\le 0.5$.
Note that although the quadratic term in $m(t)$ has a clear analogue in the full SMCT, the same is not true of this linear term. However, the approach often taken in the literature on schematic models is to add terms to $m(t)$ to access a particular arrest scenario even if terms of that order are not present in the full model (see Ref.\ \cite{goetze_les_houches}).

A further enhancement to the data-fitting capabilities of schematic MCT can be brought about by the introduction of a second correlator $\phi_A(t)$ (see e.g.\ Refs.\ \cite{sjoegren,goetze_sperl,sperl,buchalla,cummins_beta,voigtmann_propylene}). This obeys an analogous equation to Eqn.\ \ref{mcteom2}, and is coupled to the first correlator by a memory kernel
\begin{equation}
m_A(t)=v_A\phi(t)\phi_A(t)
\label{mem}
\end{equation}
The decay of $\phi_A(t)$ is thus broadly fixed by that of $\phi(t)$, whereas $\phi_A(t)$ has no effect on $\phi(t)$. Various physical motivations hve been given for the use of this extra correlator. It was initially proposed to describe the motion of a tagged particle \cite{sjoegren}, and has since been associated with the correlations of a probing variable (e.g.\ dielectric loss) and most recently with rotation-translation coupling \cite{goetze_sperl, cummins_beta}. This model can be more freely tailored to fit data \cite{voigtmann_propylene}, and may also show an intermediate-time ($\beta$ or Johari-Goldstein \cite{johari_goldstein}) decay, in addition to the usual fast (microscopic) and slow ($\alpha$) processes.

A glass transition, albeit a rather unrealistic one, is also present in the $\text{F}_1$ model, where $m(t) = v_1\phi(t)$ \cite{goetze_les_houches}. Here, $f$ does not have a jump as $v_1$ is varied but grows continuously for $v_1\ge 1$ (a {\em type A} transition), and vanishes for $v_1<1$. There is no intermediate-time plateau in the correlator. The long-time relaxation may be very slow, and a power-law decay is seen close to the transition. The model cannot produce stretched-exponential decay \cite{jacobs}.

\section{Phenomenology and theory of multi-step relaxation}\label{multistep}

As discussed above, standard mode-coupling theory predicts a characteristic two-step relaxation of density correlators near the glass transition. This is often said to be associated with the formation and breaking up of {\em cages}; each particle being trapped in a cage of its neighbors for increasingly long times as the glass transition is approached. The broad features, and many of the finer details, of this MCT scenario are seen in scattering and dielectric loss experiments on a wide range of substances (see Refs\ \cite{goetze_exp} and \cite{cummins_exp} for reviews). However, many materials show more complex relaxation patterns, with three or more decay processes (we refer to this as multi-step relaxation).

Notably, hard-sphere colloid \cite{vanmegen_prl,vanmegen_pre} experiments (which provide some of the strongest experimental support for SMCT) show a decay regime on the fluid side of the glass transition in which the deviations from SMCT fits to the slow $\alpha$ process suggest an {\em even slower} relaxation at late times. This additional slow process at long times is also hinted at in microgel experiments \cite{bartsch,bartsch2,bartsch3}, although here the $\alpha$ relaxation is less well resolved.

Glassy relaxation at {\em intermediate} times in molecular systems often shows additional processes not accounted for by SMCT when applied to density correlators only (see Ref.\ \cite{cummins_beta} for an overview). These relaxations are usually discussed in terms of the susceptibility spectrum $\chi''(\omega)$ (measured in light-scattering experiments) or the dielectric response $\epsilon''(\omega)$. The former is given by the product of the frequency $\omega$ and the Fourier cosine transform of $\phi(t)$. Its peaks correspond to the troughs of $\partial_t\phi(t)$; that is, to each decay process. The peak shapes reveal the functional form of the relaxation; in particular, exponential decay gives a symmetric (Lorentzian) peak of width $1.14$ decades, whilst stretched exponentials appear broader and asymmetric.

Intermediate-frequency relaxations are often classed into two groups. The first of these comprises positive-curvature `wings' \cite{adichtchev_glycerol,blochowicz_et_al,adichtchev_light,blochowicz_prl,wiedersich_toluene,roessler_glycerol} on the high-frequency side of the spectral peak corresponding to the $\alpha$ process. The second group includes $\beta$ {\em peaks} \cite{johari_goldstein,kudlik,olsen,ngai_dynamic,nozaki_sorbitol,roessler_beta}, appearing between the initial fast and final $\alpha$ relaxation processes. These weakly temperature-dependent \cite{johari_goldstein} peaks were initially attributed to the relaxation of intramolecular degrees of freedom; however, they were subsequently shown to occur in a wide range of substances composed of rigid molecules \cite{johari_goldstein} and postulated to be a universal feature of glassy relaxation. The $\beta$ peak is broader and weaker than the $\alpha$ peak, and tends to be rather symmetric if the frequency is plotted on a logarithmic scale \cite{johari_goldstein,goetze_beta_jncs,cummins_beta}.

The extent to which wings and peaks are manifestations of a single underlying phenomenon has been the subject of much recent debate. On the one hand, some dielectric results show the simultaneous presence of both wing and $\beta$ features \cite{kudlik,casalini_pressure}. One of these papers \cite{casalini_pressure} also finds a difference in pressure dependence between the wing and $\beta$ processes. In addition, other dielectric experiments show a connection between the excess wing and the final $\alpha$ process, either by showing that they share the same (strong) temperature dependence \cite{adichtchev_et_al}, or through fitting both by a function with a single physical motivation \cite{blochowicz_et_al,adichtchev_glycerol}.

In contrast, other experiments are able to transform a positive curvature wing into a negative curvature $\beta$ process. This can be achieved by varying the conditions of a single material, either through aging the sample \cite{lunkenheimer_wing, schneider} or changing the temperature \cite{lunkenheimer_wing_prb}. Such a transformation can also be observed by moving through a class of materials, for example by studying a sequence of alcohols with gradually decreasing hydrogen bonding \cite{doess}, or more simply by changing the composition of a binary mixture \cite{blochowicz_prl}.
The experimental situation is further complicated by subtle differences between the relaxation dynamics probed by different methods \cite{lunkenheimer_glycerol_epl,pick_tensor}. Although the excess wing has been seen in both dielectric loss \cite{lunkenheimer_wing_prb} and light scattering \cite{wiedersich_toluene,adichtchev_light} experiments, the $\beta$ peak has only (to our knowledge) been seen in dielectric measurements \cite{blochowicz_prl}.

The microscopic mechanism underlying these intermediate-frequency processes remains obscure (see Refs.\ \cite{johari_beta,ngai_dielectric,ngai_dynamic} for discussions). It is usually attributed either to rotational (and possibly translational) motion in isolated regions (see e.g.\ Ref.\ \cite{johari_beta}), or to purely rotational relaxation of all molecules (see e.g.\ Ref.\ \cite{kudlik}).

Although most discussion of $\beta$ relaxation is in the experimental literature, a number of theories containing intermediate-time processes have been proposed. The approach in which intermediate relaxations emerge most naturally is the `coupling model' \cite{ngai79,ngai_beta}. The $\beta$ process is associated with the `primitive' (i.e.\ uncoupled) relaxation time of this model, and a relation between the form of the final $\alpha$ process and the $\beta$ relaxation time is predicted, in good agreement with experiment.

As mentioned in section \ref{schematic}, intermediate-time decays are present in schematic SMCT models with a second correlator introduced via Eqn.\ \ref{mem} \cite{sperl,buchalla}. The use of an SMCT model to study such processes might be questioned \cite{cummins_beta} on the grounds that intermediate-time features often appear around the mode-coupling $T_c$, where SMCT is expected to break down (see e.g.\ Ref.\ \cite{wiedersich_toluene}). However, using this approach, excellent fits to optical Kerr effect data including wing features above the estimated $T_c$ have been obtained \cite{sperl}. By increasing the couplings, these wings were transformed into realistically broad and symmetric $\beta$ peaks, although these results correspond to lower temperatures than those considered in the experiment \cite{cang}.

The $\beta$ relaxation has also been studied \cite{zeng_kivelson_tarjus} using a simple memory equation, which (in contrast to mode-coupling) has a memory kernel with no dependence on the density correlators. Instead, a kernel consisting of the sum of two exponentials (decaying in time) is used, and a gradual separation of $\alpha$ and $\beta$ relaxation (see e.g.\ Ref.\ \cite{olsen}) as the couplings are increased is produced. The simple form of the kernel means that both these processes are purely exponential.

Note additionally that several systems (see Ref.\ \cite{fuchs_alpha} for a discussion) show multiple stretched-exponential $\alpha$ processes, but these fall outside our scope.

\section{A schematic model of the cut-off glass transition}\label{cutoff}

As discussed above, the ideal glass transition (to a perfectly arrested state with $f>0$) predicted by SMCT is not observed in real glass formers. Instead, it is cut off by relaxation processes not captured by the mode-coupling approximation. An important question is: what exactly goes wrong with SMCT in this region? 

For Newtonian systems the SMCT equations can be derived by several routes, but one of these \cite{zac} is particularly revealing, as discussed by Cates and Ramaswamy \cite{cates_ramaswamy}. This approach allows the memory function to be written formally as a sum of a `standard' contribution, which involves four-point density correlators, plus a second contribution derived from coupling of configurational to kinetic degrees of freedom. The standard contribution reduces (essentially) to the MCT form when the four-point density correlators are factored into products of two-point correlators. Cates and Ramaswamy additionally gave arguments for why the second, nonstandard, term is negligible in the glassy regime. If their arguments are accepted, then for the ideal glass transition to be avoided, it is {\em necessary} that the factorization of the four-point correlator becomes a {\em qualitatively} wrong approximation as the transition is approached.

Interestingly, in any regime dominated by localized activated processes (``hopping"), this approximation does indeed become qualitatively wrong \cite{cates_ramaswamy}. When hopping occurs in a system that would otherwise be fully arrested, the generic relaxation process in any local neighborhood involves a long wait while nothing happens. Then, after some randomly distributed time, a large local change occurs that reconfigures the density nonperturbatively (an ``instanton"). This decorrelates all powers of the density at the same instant. Indeed, if such decorrelation is complete, one has $\langle \rho^2(x,0)\rho^2(x,t)\rangle/\langle\rho^4\rangle \simeq
\langle \rho(x,0)\rho(x,t)\rangle/\langle\rho^2\rangle$: the four point correlator (at least, the one involving squared local densities) decays as the two-point correlator, not as its square. This is the extreme opposite of a Gaussian fluctuation process, whereby $\rho$ correlations decay continuously by infinitesimal increments, and the factorization of the four-point correlator as adopted in SMCT is rigorously correct \cite{zac}. 

In a general glassy relaxation process some intermediate behavior can be expected; for instance, if several instanton visits are required to achieve complete relaxation, rather than just one, a limiting approximation combining linear and quadratic terms may remain appropriate even as the ideal glass transition is approached. Additionally, while the squared-density correlator ($\langle \rho^2(x,0)\rho^2(x,t)\rangle$) does enter some early forms of MCT \cite{das_mazenko}, this in fact describes the case of a perfectly flat $S(q)$. (Equivalently, it describes a system where the direct correlation function $c(r)$, with $1-Nc(q)/V \equiv 1/S(q)$ is a $\delta$-function in real space.) Within SMCT, which addresses general forms for $S(q)$, the four point correlator entering the exact expression of \cite{zac} for the memory function is not the autocorrelator of $\rho^2$, but that of a bilinear convolution of densities with a kernel whose range is fixed by $c(r)$. This nonlocality, unless negligible on the (uncertain) length-scale of an instanton relaxation event, could also lead to a quadratic plus linear behavior of the relevant four-point correlator.

Based on these ideas, Cates and Ramaswamy \cite{cates_ramaswamy} argued that the a model of the cut-off glass transition might include a kernel in which quadratic terms are gradually replaced by linear ones, in such a manner that the system approaches the ideal glass state, acquiring some of its properties, before reverting to the liquid. A suitable choice is given by
\begin{equation}
m(t)=v_1(t/\tau_I)\phi(t)+v_2(t/\tau_I)\phi^2(t)
\label{cutoffM}
\end{equation}
where $v_1$ and $v_2$ are {\em time-dependent} coefficients and $\tau_I$ is an ``instanton timescale''. Note that $\tau_I$ need not be as long as the typical waiting time for an instanton event: it describes the time scale beyond which incremental, collective decay modes are negligible compared to instanton-mediated relaxation. Hence $\tau_I$ must lie well beyond the time for the plateau in $\phi(t)$ to be reached but in principle need not be as long as its eventual decay.

We note that this cutoff mechanism preserves an important prediction of SMCT -- the growth of the plateau height on the glass side of the ideal transition with a square-root dependence on the distance from the transition line $(\rho_c,T_c)$ (see e.g.\ Ref.\ \cite{goetze_les_houches}). Experimental data are consistent with this singularity \cite{goetze_exp}.

\begin{figure}
\includegraphics[width=\linewidth]{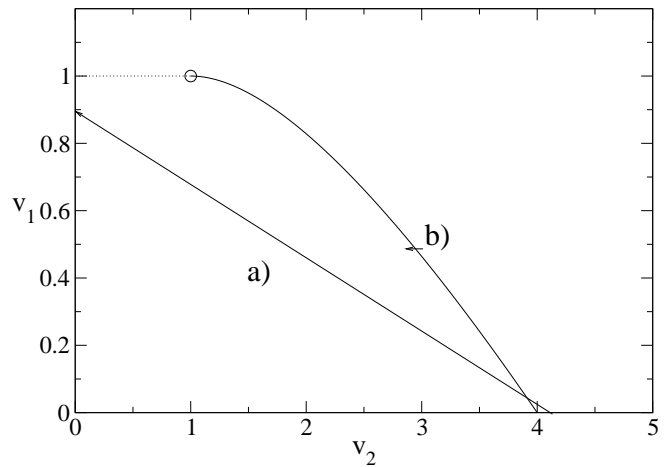}
\caption{\label{phasediagfig}
Phase diagram of the $\text{F}_{12}$ model in terms of the coupling coefficients $v_1$ and $v_2$, showing sample trajectories for our model. The dotted line shows the continuous type A transition associated with the purely linear $\text{F}_1$ model; the full line shows the discontinuous type B transition. Trajectory (a) shows gradual replacement of quadratic by linear correlation. Trajectory (b) remains close to the glass transition line at all times. We will associate (a) with colloidal systems, and (b) with molecular systems.
}
\end{figure}

This model, which we explore in detail here, thus defines trajectories in the phase diagram of the $\text{F}_{12}$ model (shown in Fig.\ \ref{phasediagfig}), with $t/\tau_I$ as the curve parameter. As the system moves along such a trajectory, it will accumulate memory of all previous states along the path. We might thus expect its behavior to differ strongly from that seen in the pure $\text{F}_{12}$ model with parameters set at either the initial {\em or the final} points on such a trajectory.

As is traditional with schematic models, we put constraints on the model parameters (or here, the trajectory) to ensure that
the desired physics is recovered. To avoid permanent arrest, we  must choose $v_2(t\to\infty/\tau_I)<4$ and $v_1(t\to\infty/\tau_I)<2v_2^{1/2}(t\to\infty)-v_2(t\to\infty)$ \cite{goetze_les_houches}, so that the system moves into the liquid state at long times $t\gg \tau_I$ (see Fig.\ \ref{phasediagfig}). To have a cut-off glass transition, rather than simply a crossover from one set of liquid-like parameters to another, we must set $v_1(t=0)>v_2^{1/2}(t=0)-v_2(t=0)$ which ensures the system begins its trajectory within the ideal glass (although we will also consider trajectories starting close to the ideal glass transition on the liquid side). We of course require the model to start with {\em some} quadratic correlations, so that $v_2(t=0)>0$. We also demand $v_1(t\to\infty)>0$, so that {\em some} linear correlation remains at long times. 

Within the above constraints, there is still considerable freedom in the choice of initial and final states; in the path taken between them; and in the time spent on different sections of that path. Thus we may begin with an $\text{F}_2$ or $\text{F}_{12}$ model of the ideal glass, and leave differing amounts of linear and quadratic correlation in the final state. For the present, we shall restrict ourselves to straight paths in the $(v_1, v_2)$ plane of the general form
\begin{equation}
v_{1,2}^f + (v_{1,2}^i-v_{1,2}^f)f(t/\tau_I)
\label{decay}
\end{equation}
where $f(t/\tau_I)$ is a crossover function satisfying $f(0)=1$ and $f(t\to\infty/\tau_I)=0$, and $i$ and $f$ superscripts denote initial and final states respectively. Sample trajectories are shown in Fig.\ \ref{phasediagfig}. Trajectory (a) remains closest to the original theoretical arguments \cite{cates_ramaswamy}, with all quadratic contributions gradually replaced by linear. Trajectory (b) remains close to the ideal glass transition line at all times, crossing from just inside the glass to just inside the liquid: the motivation for this will be discussed later.

\section{Numerical procedure}\label{numerics}

We solve the cut-off schematic MCT equation of motion (Eqns.\ \ref{mcteom2} and \ref{cutoffM}) in the time domain using an algorithm introduced and discussed in detail in Ref.\ \cite{fuchs_alpha}. This procedure is based on the fact that $\phi(t)$ is fixed by $\phi(t'<t)$ and $m(t'<t)$. It contains two important technical steps: the separation of slow and fast variations in the memory integral, and the use of decimation. The decimation allows the equation of motion to be solved on progressively coarser grids as $t$ is increased: any fast decay occurs at short times and requires fine resolution, while slow decay occurs at long times and can be calculated on a coarser mesh. First, the equation is solved for $N$ points with step size $\delta t$ over an interval $T$. The solution is then transferred to a grid of $N/2$ points with step size $2\delta t$ by taking the weighted average of groups of neighboring points, and Eqns.\ \ref{mcteom2} and \ref{cutoffM} are then solved up to $2T$. The procedure is repeated until the final relaxation has been resolved. At short times, a Taylor series expansion for $\phi(t)$ is used. The time derivatives are discretized using interpolation polynomials (see e.g. Ref.\ \cite{abramowitz}). At each timestep, $\phi(t)$ is recalculated until it converges to a relative accuracy of $10^{-9}$, a maximum of $1000$ iterations being allowed. We use a grid of $60$ to $100$ blocks of $256$ points each. The stepsize in the first block is $10^{-9}$, increasing to $2\times 10^{-9}$ in the second, and so forth. The first $50$ points are calculated using the short-time expansion.
All Fourier-transformed quantities are calculated using a simplified Filon algorithm with linear (rather than quadratic) interpolation \cite{tuck}. Here, we use up to $24$ blocks of $180$ points, each block corresponding to a decade in frequency.

A representative sample of the results for $\phi(t)$ was then reproduced using a (much slower) iterative procedure based on the Laplace transform of Eqns.\ \ref{mcteom2} and \ref{cutoffM} \cite{goetze_scaling}, with the same numerical parameters as the Filon algorithm. These checks, combined with those made by previous authors who have used and developed such algorithms, inspires confidence that the numerical results reported below are accurate solutions of the governing equations.

\section{Results}\label{results}

We divide our results according to whether the final point on the trajectory has or has not a nonzero value of the quadratic coupling
coefficient, $v_2(t\to\infty)$.

\subsection{Crossover to pure linear coupling}\label{purelinear}

To begin, we concentrate on trajectories (such as (a) in Fig.\ \ref{phasediagfig}) where quadratic coupling is entirely removed at long times. These maintain the closest connection with the original theoretical motivation of the model \cite{cates_ramaswamy}. 

Several features of Eqn.\ \ref{decay} (starting and finishing points, crossover function, decay time) may, in principle, be freely adjusted. The microscopic frequency $\Omega$ and the damping term $\nu$ are also free; however, $\Omega$ serves simply to fix the unit of time and $\nu$ only has significant influence on the short-time dynamics \cite{goetze_alpha}. In the following, we choose $\Omega=1$ and $\nu=10$ (to avoid pronounced oscillations at short times). A selection of results were checked with $\nu=20$; as expected, this only appreciably changed the initial relaxation.

We now determine in what way, and how strongly, adjusting Eqn.\ \ref{decay} affects the calculated form of $\phi(t)$. Here, and in the following, we will often show our results as plots of $-\ln(\phi(t)/\phi_p)$ vs.\ $t$ on a log--log scale; here $\phi_p$ is the value of $\phi(t)$ on the plateau of the decay. This representation \cite{goetze_scaling} is designed to isolate the final relaxation away from the plateau. Stretched exponential relaxation ($\propto \exp(-(t/\tau)^\beta)$, with $\tau$ some characteristic timescale) will appear as a straight line, with slope equal to the stretching exponent $\beta$. The standard $\phi(t)$ versus $\log t$ plots will be shown in insets.

Some considerations concerning the structure of our model allow us to restrict the choice of starting point. We wish to study the crossover {\em from} a model with significant quadratic correlation: this rules out starting near the type A transition where linear correlation is dominant. We also follow early work on schematic models \cite{goetze_scaling} in starting close to (within $\pm 0.2$ of) the type B transition. Starting at higher couplings would introduce a substantial region of decay determined entirely by the arbitrary crossover function $f(t/\tau_I)$ and the properties of the SMCT model deep in the (unrealistic) ideal glass. Furthermore, the effects of such a decay would propagate through to all later times. In this subsection (\ref{purelinear}), we will specialize further to paths with $v_1^i=0$. This is the most literal implementation of our theoretical arguments. In addition, although we have verified that the results for trajectories with $v_1^f\ne0$ are broadly similar, the crossover phenomena we wish to discuss emerge most clearly when the initial state contains no linear correlation.

\begin{figure}
\includegraphics[width=\linewidth]{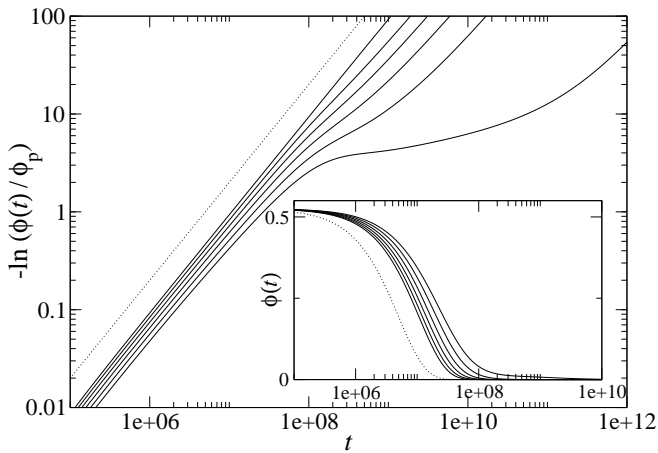}
\caption{\label{endfig}
Long-time decay for a range of end points. In all cases, the start point is $v_1^i=0, v_2^i=4.01)$. From top to bottom, $(v_1^f=0, v_2^f=0)$, $(0.2,0)$, $0.4,0)$, $(0.6,0)$, $(0.8,0)$, $(0.99,0)$. $f=\exp(-t/\tau_I)$; $\tau_I=10^8$. Dotted lines show exponential decays.
}
\end{figure}

The system shows very strong sensitivity to the finishing point of the trajectory. Figure \ref{endfig} shows decay curves for fixed starting point $(v_1^i=0,v_2^i=4.01)$, fixed $v_2^f=0$, $v_1^f$ varying from $0$ to $0.8$ in steps of $0.2$ and an additional value of $v_1^f=0.99$ (just within the liquid). In all cases, $\tau_I=10^8$. For $v_1^f=0$ (top curve), the final decay is purely exponential. However, as $v_1^f$ increases, a region of slower-than-exponential decay develops at $t\sim\tau_I$. For larger $v_1^f$, this decay becomes extremely slow, as may also be seen from the inset to Fig.\ \ref{endfig}. This behavior may be attributed to the growth of linear correlations: having moved away from one glass transition, the system approaches another (that of the $\text{F}_1$ model), and its dynamics acquire an additional slow contribution. At very long times, the decay becomes exponential. This is a generic feature of all MCT models with discrete wave-vectors \cite{goetze_alpha} (which naturally includes single-correlator models).

The relative importance of the trajectory endpoint in determining the decay of $\phi(t)$ might be expected: for any reasonable choice of crossover function, the system will spend an infinite amount of time on the final section of the path, however small this is taken to be.

\begin{figure}
\includegraphics[width=\linewidth]{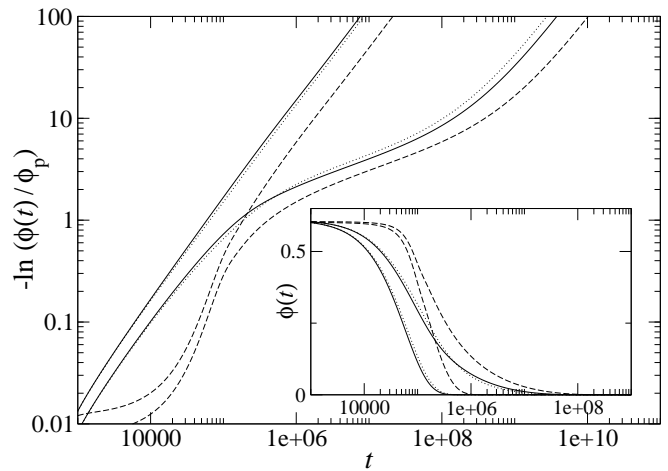}
\caption{\label{crossover}
Range of crossover functions $f(t/\tau_I)$. All start at $(v_1^i=0,v_2^i=4.2)$ and finish at $v_2^f=0$. The upper trajectory has $v_1^f=0.05$ and the lower has $v_1^f=0.95$. Each trajectory is plotted for an exponential crossover function (solid line), a power law ($\propto 1/(1+t/\tau_I)$) (dotted line) and a very sharp logistic law ($\propto 1/(1+\exp(5(t-\tau_I)/\tau_I))$) (dashed). In all cases, $\tau_I=10^5$. The dotted lines on the left of the plots show exponential decay.
}
\end{figure}

We now check the sensitivity of the system to the choice of crossover function $f(t/\tau_I)$. Figure \ref{crossover} shows two trajectories: both start at $(v_1^i=0,v_2^i=4.2)$ and finish at $v_2^f=0$. The upper trajectory has $v_1^f=0.05$ (very far from the type A transition), and the lower has $v_1^f=0.95$ (close to the type A transition). Each trajectory is plotted for an exponential crossover function (solid line), a power law ($\propto 1/(1+t/\tau_I)$) (dotted line) and a very sharp logistic law ($\propto 1/(1+\exp(5(t-\tau_I)/\tau_I))$) (dashed). In all cases, $\tau_I=10^5$. At longer times ($\tau\ge 10^6$), the influence of $f(t/\tau_I)$ is not strong, all trajectories lie close to parallel, and the choice of final coefficients plays the most important role in fixing the behavior of $\phi(t)$. This is reassuring, given that theory provides no clear arguments as to how $f(t/\tau_I)$ should be chosen. The crossing of the exponential and power-law lines for $v_1^f=0.95$ can be understood by realising that while the exponential gives a quicker initial decay, it also approaches the type A transition more rapidly, thereby acquiring an additional slow contribution at earlier times than the power-law trajectory.

However, the situation is different at times $\sim\tau_I$. In this region, the decay curves fall into groups according to the choice of $f(t/\tau_I)$. This means that the detailed form of the initial decay from the plateau will be sensitive to the crossover function: we return to this point later. Unless otherwise stated, we use an exponential crossover in the following.

\begin{figure}
\includegraphics[width=\linewidth]{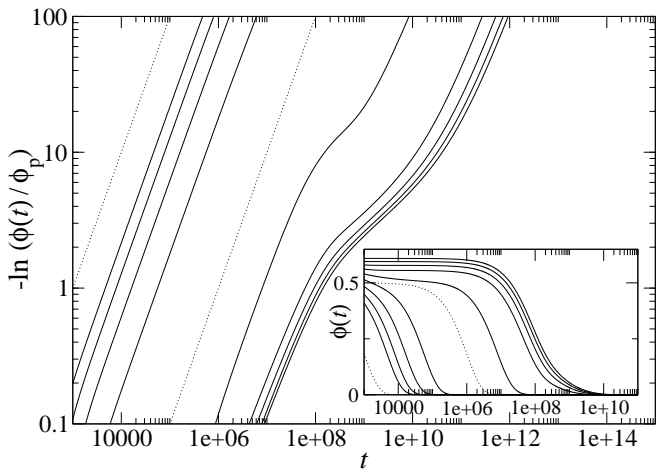}
\caption{\label{tau1e8}
Range of starting points on either side of the ideal glass transition, moving from $(v_2^i=3.8)$ to $(v_2^i=4.2)$ in steps of $0.05$. In all cases, $v_1^i=0$, $(v_1^f=0.9,v_2^f=0)$ and $\tau_I=10^8$. For this large value of $\tau_I$, trajectories starting on the liquid side of the line are only weakly affected by the cutoff; those on the glass side show extra slower-than-exponential relaxation at long times. Dotted lines show exponential decay.
}
\end{figure}

The final free quantity in the model is the decay time $\tau_I$. On the glass side of the ideal transition, this directly sets the timescale of the final relaxation (see the rightmost four trajectories in figure \ref{tau1e8}, where $\tau_I=10^8$), and varying $\tau_I$ leaving the other model parameters fixed simply serves to shift the final decay along the time axis. Close to the transition on the liquid side, the role of $\tau_I$ depends on how close it is to the timescale of the $\alpha$ relaxation in the unmodified schematic model. In figure \ref{tau1e8}, we plot a series of trajectories, all with a rather large decay time ($\tau_I=10^8$) and all finishing at $(v_1^f=0.9,v_2^f=0)$ (close to the type A transition). The starting point is varied from $(v_1^i=0,v_2^i=3.8)$ to $(v_1^i=0,v_2^i=4.2)$ in steps of $0.05$, crossing the ideal glass transition at $v_2=4$. In this case, $\tau_I$ is too long to interfere significantly with the relaxation in the approach to the ideal transition from the liquid side, and the $\phi(t)$ here are very close to those of the unmodified $\text{F}_2$ model. In particular, they do not show the extra slow process at long times discussed above, which only appears on the glass side of the ideal transition.

\begin{figure}
\includegraphics[width=\linewidth]{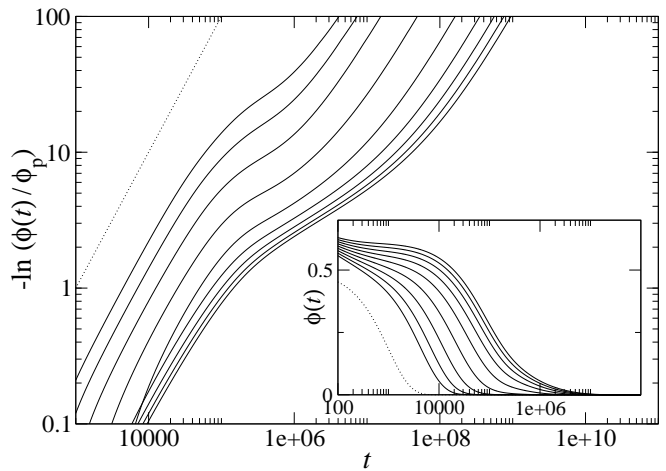}
\caption{\label{tau1e5}
Range of starting points on either side of the ideal glass transition, moving from $(v_2^i=3.8)$ to $(v_2^i=4.2)$ in steps of $0.05$. In all cases, $v_1^i=0$, $(v_1^f=0.9,v_2^f=0)$ and $\tau_I=10^5$. For this smaller value of $\tau_I$, trajectories on both sides of the SMCT transition line show extra slower-than-exponential relaxation at long times. Dotted lines show exponential decay.
}
\end{figure}

If we choose a shorter decay time $\tau_I=10^5$ (still much greater than the time required to reach the plateau) (figure \ref{tau1e5}), the $\alpha$ relaxation close to the ideal transition on the {\em liquid} side may also be affected by the cutoff, acquiring an additional slow process at long times as described above. Again, the timescale of the relaxation on the glass side of the ideal transition is directly set by $\tau_I$.

All these trajectories retain no quadratic coupling as $t\to\infty$. All such trajectories starting within the ideal glass (or close to it and with a sufficiently short $\tau_I$) and finishing with $v_1^f\ne 0$ display certain generic long-time properties: an initial exponential decay from the plateau (inherited from the $\text{F}_2$ model) followed by a rather rapid crossover to a slower-than-exponential decay. This produces a characteristic kink in the $-\ln(\phi(t)/\phi_p)$ vs.\ $t$ plots. Such behavior is straightforward to rationalize in terms of crossover between the starting and finishing models. 

Also, it is not unlike the behavior seen in light scattering experiments on colloidal systems \cite{vanmegen_prl,vanmegen_pre}.
Note that these colloid data are usually fitted to SMCT with shifted parameters, chosen to bring the ideal and the actual glass transition into register. This gives reasonable agreement, but, as mentioned before, the deviations suggest an unaccounted slow process beyond the $\alpha$ relaxation time. (Note however that the number of data points at these long times is often limited.) One alternative viewpoint is to assert that for densities between the ideal and the actual glass transition $\rho_c\le\rho\le\rho_g$, the observed $\alpha$ relaxation is the result of an instanton-induced cut-off and should be calculated accordingly, not by shifting parameters in SMCT. For suitably chosen trajectories within the schematic model, this could account in principle for the deviations.

However, there is considerable experimental evidence to support the original identification of the arrest in colloidal fluids with the SMCT transition \cite{vanmegen_prl,vanmegen_pre}. In these materials, the absolute values of the viscosity are very high, and the system appears arrested even when the slow relaxation covers only around five decades after the microscopic relaxation (in contrast to more than ten decades in molecular glasses). This means that the system may become glassy at (relatively speaking) lower couplings, and the experimentally-observed arrest may occur close to $\rho_c$.

Furthermore, the scaling properties of $\phi(t)$ (see e.g. Ref.\ \cite{goetze_les_houches}) are verified on this assumption (i.e.\ $\rho_g=\rho_c$). In addition, the plateau becomes clearer at $\rho_c$ and its subsequent growth is consistent with the MCT cusp singularity \cite{goetze_les_houches,goetze_exp}). It may therefore be that a scenario such as that shown in figure \ref{tau1e5} is more appropriate. Here, $\tau_I$ is shorter, and the cutoff interferes with the $\alpha$ relaxation close to the transition on the liquid side, producing a slow final relaxation. It should be noted that the above figures should not be interpreted literally as sets of results for a colloidal system at different densities. For each figure, the relaxation time $\tau_I$ is fixed, but in real systems some density dependence would be expected.

\subsection{Crossover to mixed coupling}\label{mixed}

We now consider another class of trajectories (such as (b) in Fig.\ \ref{phasediagfig}). These also begin on the glass side of the type B transition and cross over to the liquid. However, we now choose the final point to lie fairly close to the liquid-glass transition line, so that both quadratic and linear coupling remain at long times. Though moving away somewhat from the arguments originally presented in Ref.\ \cite{cates_ramaswamy}, a rationale for retaining some quadratic coupling was given in Section \ref{cutoff} above. Additionally, the connection between the full and schematic memory kernels is not completely clear, so that an endpoint with a reduced but non-zero quadratic coupling might in fact be the best schematic representation of a full $q$-dependent model with purely linear coupling (rather as the $\text{F}_{12}$ model is, by common consent, the best schematic representation of the full model with purely quadratic coupling). One might further argue that, as soon as instanton processes have made the system fluid, the dynamics should again be determined by SMCT. Also, to our knowledge, the `slower-than-$\alpha$' process discussed above does not occur in molecular glasses, and the low-frequency side of the $\alpha$ peak is well described by SMCT both above and below $T_c$ . It then seems reasonable to consider trajectories finishing close to the ideal transition line with the intention of preserving this agreement. To obtain a stretched exponential final decay, we use $v_1^i\ne0$, corresponding to an $\text{F}_{12}$ model.

In atomic and molecular (as opposed to colloidal) systems, glassy relaxation is often discussed in terms of the spectrum $\chi''(\omega)$, defined in Section \ref{multistep}, and we will adopt this representation in the following.

\begin{figure}
\includegraphics[width=\linewidth]{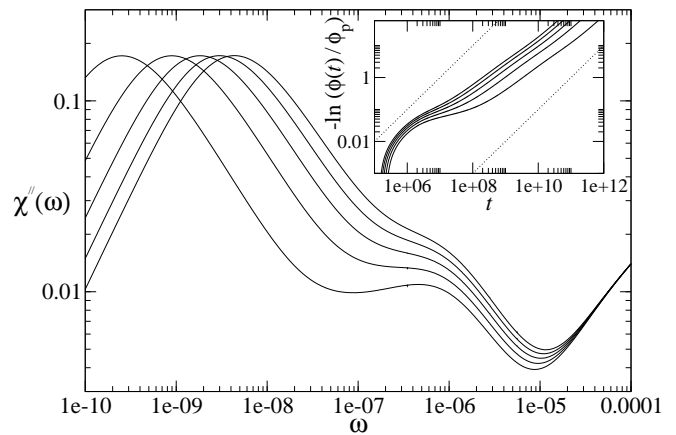}
\caption{\label{beta}
Range of loss spectra showing an intermediate-frequency decay process. These results should be compared with the dielectric data of Refs \cite{olsen, nozaki_sorbitol,nozaki_water-sorbitol,sudo_glycerol-water}. Inset shows the same trajectories plotted as $-\ln(\phi(t)/\phi_p)$ vs.\ $t$, with the dotted lines showing exponential decay. The stretched-exponential nature of the final relaxation is clearly visible. In all cases, $v_1^i=v_1^f=0.5$, $v_2^i=2.9242$ and $f=\exp(-t/\tau_I)$ with $\tau_I=10^6$. $v_2^f$ is varied from $2.9042$ to $2.9122$ in steps of $0.002$ (running from right to left in the main figure and left to right in the inset). These trajectories start and finish rather close to the glass transition line.
}
\end{figure}

Sample $\chi''(\omega)$ for short trajectories starting and finishing close to the glass transition line are shown in Fig.\ \ref{beta}, and the corresponding $\ln (\phi(t)/\phi_p)$ plots are shown in the inset. The final slow decay process associated with the dominance of linear coupling is now absent, leaving a standard stretched exponential $\alpha$ relaxation (see inset to Fig.\ \ref{beta}), which moves out to longer times as the end of the trajectory is moved closer to the ideal glass line. However, a weaker process now emerges at intermediate times.

Again, it is straightforward to rationalize the relaxation behavior in terms of the contributions of different stages of the trajectory. Thus $\phi(t)$ first decays to a plateau set by the glass side of the ideal transition. At times $t\sim\tau_I$, it crosses to a conventional $\text{F}_{12}$ viscous liquid state, which now dominates the long-time dynamics and leads to a standard final $\alpha$ relaxation. To understand the form of the intermediate-time relaxation, we note that the plateau height in the viscous liquid is constant within MCT, but grows (with a square root dependence on the distance from the transition line) in the glass \cite{goetze_rev}. This means that, on crossing the ideal glass line, $\phi(t)$ falls from the glass plateau value to that of the liquid, leading to a decay at times $t\sim\tau_I$. This mimics the $\beta$ relaxation discussed earlier and gives a good qualitative agreement with dielectric loss experiments on intermediate-fragility liquids with a degree of hydrogen bonding (compare Fig.\ \ref{beta} with, for example \cite{olsen,nozaki_sorbitol,nozaki_water-sorbitol,sudo_glycerol-water}), with a weak $\beta$-like relaxation that persists, remaining at a constant frequency, as the $\alpha$ relaxation moves to lower frequencies. 

\begin{figure}
\includegraphics[width=\linewidth]{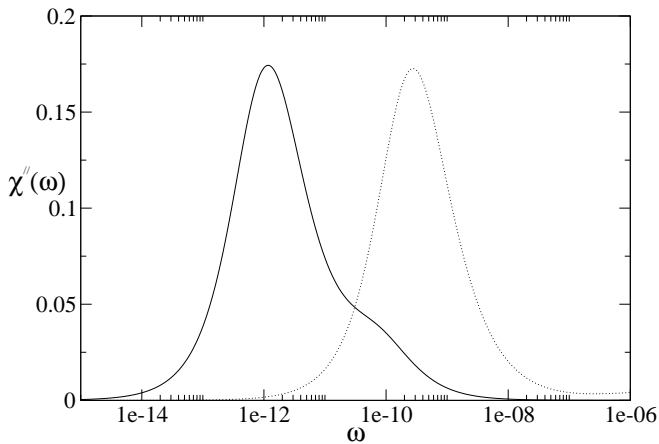}
\caption{\label{winglarge}
Spectra showing the sudden appearance of the intermediate-frequency relaxation as the starting point of the trajectory is moved to the glass side of the glass transition line. The mode-coupling transition can thus be associated with a qualitative change in the dynamics of the system. $v_1^i=v_1^f=0.5$, $v_2^f=2.864$ and $\tau_I=10^6$. $v_2^i=2.914$ (full line) and $v_2^i=2.864$ (dotted line). For this $v_1$, the glass transition occurs at $v_2\approx 2.91421$.
}
\end{figure}

The association of the intermediate-time process with the ideal glass transition line is demonstrated in Fig.\ \ref{winglarge}. Here, we plot the loss spectra of two trajectories, both finishing at the same state point in the viscous liquid. One begins in the glass; the other begins in the liquid very close to the transition line. The former trajectory shows a clear `$\beta$' process; in the latter, it is completely absent. The intermediate decay thus emerges as a precursor to the final relaxation, present only below $T_c$.

A similar connection between the ideal SMCT glass transition and a qualitative change in dynamics (often involving the emergence of an intermediate-time relaxation) has been made many times in the experimental and simulation literature \cite{debenedetti_rev,sastry,roessler_beta,adichtchev_et_al,adichtchev_light,blochowicz_et_al,wiedersich_toluene}. However, we must state that the relation between experimental variables (density and temperature) and the trajectories chosen within our model -- particularly the values of the couplings in the final state -- remains unclear. We might associate increasing coupling with increasing density and decreasing temperature, but it is not obvious how to be more precise than this.

The fact that our intermediate-time process arises from the crossing of the ideal glass line implies that this relaxation should be present for all trajectories finishing on the liquid side of the line, including those that end only just within the liquid, so that the subsequent $\alpha$ decay is pushed out to unobservably long times. This type of model has interesting properties which we now discuss, although these become increasingly remote from the original conception of the instanton-mediated cutoff. Specifically, we are now talking about a model where the instanton relaxation only just manages to carry the parameters into the fluid phase, such that the resulting $\alpha$ relaxation remains much slower than the instanton time itself. 
This behavior could perhaps arise at temperatures well below $T_c$, but if so, decreasing temperature further should create a trajectory where even the final state lies within the ideal glass.
This would cause not only the final divergence of the $\alpha$ relaxation time (which could be too long to measure far before that) but also the divergence of the $\beta$-like relaxation time.

\begin{figure}
\includegraphics[width=\linewidth]{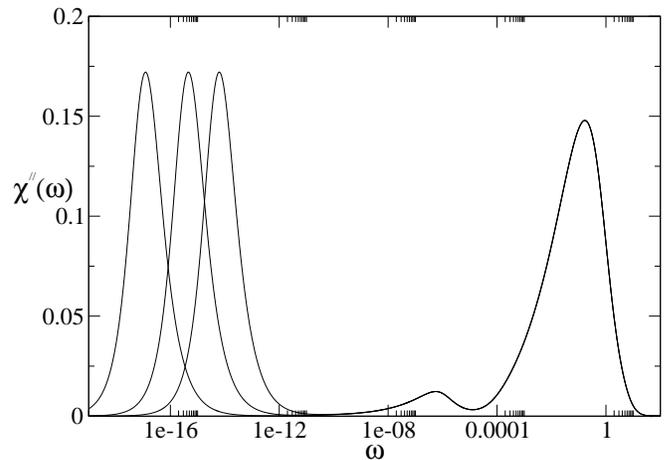}
\caption{\label{slow_alpha}
Spectra showing the persistence of the intermediate-frequency process as the final decay is moved to very low frequencies. $v_1^i=v_1^f=0.5$, $v_2^i=2.9242$ and $\tau_I=10^6$. $v_2^f$ is moved very close to the transition line: the values used are $2.9142$, $2.91421$ and $2.914213$.
}
\end{figure}

Until this point is reached, however, the $\beta$-like relaxation arising from the instanton cutoff mechanism should persist at all temperatures below the mode-coupling transition, even when the system is arrested (no visible $\alpha$ decay) on experimental timescales.  
To demonstrate this, we plot (in fig.\ \ref{slow_alpha}) the loss spectrum for three trajectories ending very close to the glass transition line on the liquid side. The intermediate peak persists even when the $\alpha$ peak occurs at frequencies twelve orders of magnitude lower.

As in the case of the earlier trajectories with $v_1^f=0$, the qualitative form of our results is largely determined by the couplings in the final state of the system, provided the trajectory begins in the glass phase. However, there are a number of differences between the two classes of trajectory in the way the various adjustable parameters affect the relaxation.

Firstly, the role of the decay time $\tau_I$ can be different in the shorter trajectories. Figure \ref{slow_alpha} demonstrates this: here, $\tau_I$ sets the intermediate relaxation time, but the difference between this time and that of the $\alpha$ relaxation is determined by the distance between the finishing point and the glass transition line. This is in contrast with the $v_1^f=0$ case, where the influence of the transition region is much weaker and the only way of significantly changing the timescale of the final relaxation on the glassy side of the ideal transition is through adjustment of $\tau_I$.

\begin{figure}
\includegraphics[width=\linewidth]{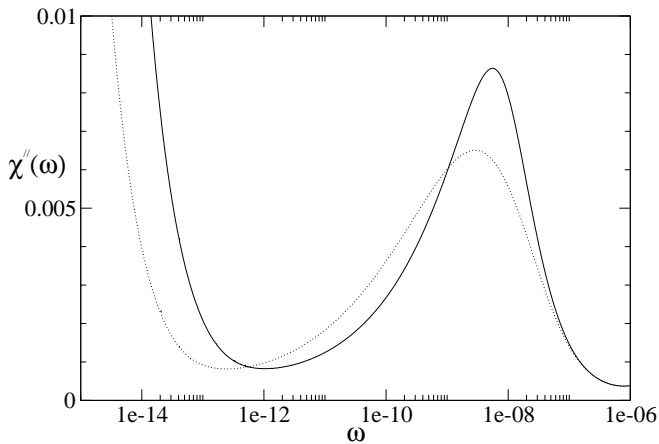}
\caption{\label{expvpower}
Sections of spectra for exponential (full line, narrower peak) and power-law (dotted line, broader peak) crossover functions $f(t/\tau_I)$. In both cases, $v_1^i=v_1^f=0.5$, $v_2^i=2.92421355$, $v_2^f=2.9142$, and $\tau_I=10^8$. These trajectories are chosen to give a well-separated intermediate-frequency process, and so finish very close to the glass transition line.
}
\end{figure}

The second point concerns the crossover function $f(t/\tau_I)$. As discussed earlier, this strongly affects the relaxation at times $\sim\tau_I$: the timescale (by construction) of our intermediate-time process. This is demonstrated in Fig.\ \ref{expvpower}: the use of a slower crossover function (here, a power law $\propto 1/(1+t/\tau_I)$) broadens the intermediate peak. A theory of $f(t/\tau_I)$ would be necessary in order to make detailed comparisons of a cut-off model with experimental data: at present, our rather narrow and asymmetric peaks are far from the broad and (usually) symmetric features seen in experiment.

We note that it is possible to obtain qualitatively similar relaxation patterns (weak intermediate process followed by stronger $\alpha$ decay) regardless of the angle of approach to the glass transition line, even if the strength of the {\em linear}, rather than the quadratic coupling is reduced. Although this moves even further away from our original theoretical motivation, it reinforces the view that the intermediate-time process is broadly insensitive to the details of the trajectory.

\begin{figure}
\includegraphics[width=\linewidth]{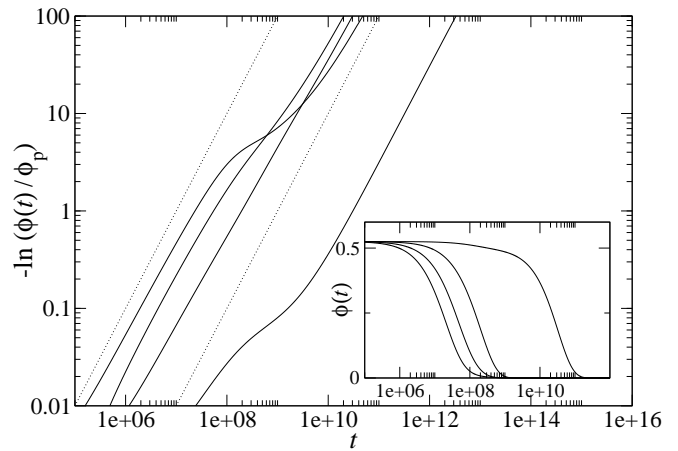}
\caption{\label{diag4.01}
Range of endpoints lying along a diagonal line in the phase diagram running from $(v_1^f=0,v_2^f=4)$ to $(v_1^f=1,v_2^f=0)$. In all cases, starting point is $(v_1^i=0,v_2^i=4.01$ and $\tau_I=10^8$. Endpoints shown are (moving upwards) $(v_1^f=0.0025,v_2^f=3.99)$, $(0.1125,3.5)$, $(0.5,2)$ and $(0.9,0)$. Dotted lines show exponential decay.
}
\end{figure}

Finally, mindful of the fact that we have focused in this section exclusively on trajectories finishing close to the type B transition, we examine briefly the behavior as the final quadratic coupling is reduced back towards zero. Figure \ref{diag4.01} shows a series of decays with starting point $(v_1^i=0,v_2^i=4.01)$, $\tau_I=10^8$ and endpoints along a diagonal line running across liquid region of the phase diagram from $(v_1=0,v_2=4)$ to $(v_1=1,v_2=0)$. The endpoints shown are $(0.0025,3.99)$ (lowest curve), $(0.1125,3.5)$, $(0.5,2)$ and $(0.9,0)$ (highest curve). Although the $(0.0025,3.99)$ trajectory shows a marked $\beta$-like process, such a relaxation is completely absent from the $(0.1125,3.5)$ trajectory, which (despite finishing rather close to the type B transition) gives a straight line in the logarithmic plot corresponding to a slightly stretched exponential decay. The appearance of the intermediate-time process thus requires the trajectory to finish very close to the type B transition (although this requirement is less strict for higher initial coupling constants e.g.\ $v_2^i\approx 4.2$). For the `central' endpoint at $(0.5,2)$, the influence of the type A transition is already noticeable, and the additional slow process at long times has started to appear. This process is very obvious in the trajectory finishing at $(0.9,0)$. We have found no trajectories showing {\em both} the final slow process and a $\beta$-like relaxation.

\section{Conclusions}\label{conclusions}

We have studied a simple schematic mode-coupling equation with cutoff, based on the well-studied $\text{F}_{12}$ model but with coupling constants in the memory kernel $m(t)$ that themselves have explicit $t$-dependence. The quadratic coefficient in the expression relating the memory kernel to the density-density correlator gradually decays, cutting off the ideal glass transition at long times. This leads to two new decay scenarios, both of which qualitatively resemble experimental data in different regimes.

In the first of these scenarios, the decay of the quadratic coupling is complete, leaving a purely linear memory kernel at long times. This is close to the conception of Cates and Ramaswamy concerning the likely form of the memory function in the hopping-dominated regime \cite{cates_ramaswamy}. For systems that standard MCT would predict to lie within or close to the ideal glass, but where a long time $\alpha$ decay is nonetheless observable on experimental timescales, this can result in decay curves showing an addition slow relaxation feature at long times. Although the experimental data is not conclusive, such features are hinted at in several of the earliest papers on hard-sphere colloidal glasses \cite{vanmegen_prl,vanmegen_pre,bartsch,bartsch2,bartsch3}. There, they appeared as an upward deviation between $\phi_q(t)$ and the relaxation predicted by a SMCT calculation with parameters shifted somewhat, so as to make the $\alpha$ time scale finite. 

In the second scenario, the decay of the quadratic coupling again carries one through the ideal glass transition line, but saturates at values not far from that line so that the final $\alpha$ process does indeed resemble the MCT prediction with shifted parameters. Theoretical motivation for this type of model is less clear, but the results are intriguing nonetheless. Specifically, a weaker relaxation process appears at intermediate times, as a precursor of the final $\alpha$ relaxation. This may be connected with the $\beta$ relaxation and/or the ``excess wing'' seen in many molecular glass-formers (see, for example Refs.\ \cite{olsen,nozaki_sorbitol,nozaki_water-sorbitol,sudo_glycerol-water}), whose appearance is clearly connected to an underlying ideal mode-coupling glass transition (cf.\ Refs.\ \cite{debenedetti_rev,sastry,roessler_beta,adichtchev_et_al,adichtchev_light,blochowicz_et_al,wiedersich_toluene}. Our schematic model predicts the $\beta$-like relaxation to persist even as the final $\alpha$ relaxation time becomes unmeasurably long. 

This second scenario requires that significant quadratic coupling remains within the memory function even as $t\to\infty$. 
If this does happen, one might expect the strength of this coupling to increase as one moves further into the glass, just as the initial ($t=0$) quadratic coupling does (unless there is a balancing compensation in the strength of the decay). The result would be that at some low enough temperature the $\alpha$ relaxation time actually diverges; if so, that of the $\beta$-like process does also. This is because the intermediate timescale corresponds to that on which the parameter decay takes the system across the ideal glass transition, and a divergent $\alpha$ relaxation time signifies that this line is never crossed.

Although the case for residual quadratic coupling within the schematic model remains somewhat unclear, several arguments in favour were presented in Sections \ref{decay} and \ref{mixed}. An additional, speculative idea is that a stronger residual coupling would result from stronger static correlations in the system, perhaps due to intermolecular attractions (e.g.\ hydrogen bonding), as often occur in molecular glasses. In that case, one might find that in hard-sphere glasses, where intermolecular attraction is absent, the reduction of quadratic coupling becomes the slowest process in the system, and determines its long-time behavior (scenario 1 above). In the molecular glasses, where attraction and bonding may enhance arrest, it is only the second slowest, and manifests itself as an intermediate relaxation (scenario 2).

The cut-off procedure studied here applies regardless of the microscopic dynamics of the system (see discussions in Refs.\ \cite{szamel_epl,mayer}). In particular, we have introduced no coupling to density {\em currents} \cite{goetze_emct,das_mazenko}, which are present in Newtonian but not Brownian systems. In fact, the experimental results on colloidal systems \cite{vanmegen_prl,vanmegen_pre} that suggest an extra decay process at long times (as predicted by our model) could be taken to indicate the presence of non-MCT relaxation processes in simple systems with Brownian dynamics and hard-sphere interactions. 

A criticism that might be made of our work is that the memory kernel of our model now contains an extra non-microscopic relaxation process introduced by hand (in contrast to SMCT) and so the fact that we predict an extra relaxation process in $\phi(t)$ is hardly surprising. However, we believe that our approach has sufficient theoretical motivation \cite{cates_ramaswamy}, and captures enough interesting features of glassy relaxation, for its study to be justified. We also have no method of calculating the relaxation time $\tau_I$, which, for example, precludes a study of the connection between $\alpha$ and $\beta$ relaxation times. However, since $\tau_I$ is connected with the breakdown of MCT, we do not expect it to be calculable by MCT-like methods; to whatever extent instanton dynamics (hopping) is actually involved in the $\beta$ process, agreement with models that omit it must be fortuitous.

Our model also has several limitations inherited directly from the $\text{F}_{12}$ model: these restrictions in plateau height, in the form of the $\alpha$ decay, and in the connections between the two. The form of the $\text{F}_{12}$ phase diagram (Fig.\ \ref{phasediagfig}) is also rather specific to this model \cite{goetze_les_houches}. However, we believe that the broad features of our results should be independent of the details of Fig.\ \ref{phasediagfig}. The appearance of the `slower-than-$\alpha$' process in our first relaxation scenario depends only on the presence of significant linear correlations at long times, not on the form of the path taken across the phase diagram. As discussed towards the end of Section \ref{results}, the intermediate process is produced whenever the system crosses the type B transition line and remains sufficiently close to it, regardless of the details of the trajectory.

Given the sensitivity of the predicted intermediate-time relaxation to the crossover function $f(t/\tau_I)$, theoretical work on the likely form of this would be useful. This might allow connections to be made between our approach and facilitated dynamics models \cite{garrahan_chandler,geissler_reichman}, which provide information on the behavior of strongly and weakly correlated regions as a system relaxes. 

Nonetheless, the lack of microscopic detail in our model might be seen as an advantage. For instance, it provides a simple mechanism by which a single generic correlator can acquire an intermediate-time relaxation (triggered by the ideal MCT transition and persisting even if the system is arrested on experimental timescales), without appealing to any detailed properties of the material or a coupling to a probing variable. This is intriguing given recent experimental results \cite{roesner_metal} suggesting an intermediate-time process in a metallic glass. These experiments are not based on scattering or dielectric loss, but instead involve mechanical measurements of the shear modulus. Furthermore, the atoms of these materials (at least with respect to their slow dynamics at high densities) are expected to behave as spheres with almost purely metallic interactions \cite{roesner_metal}. Since our model predicts a separate intermediate-time process in the main $\phi(t)$, rather than in a second probing correlator \cite{buchalla,sperl}, it also implies that intermediate-time decay should be observable in simulations, where $\phi(t)$ is measured directly. However, resolving an intermediate-time process in simulations may prove technically difficult: these decays are often small in amplitude and the intermediate-time regime may be affected by oscillations caused by the short-time dynamics (although this might be avoided by the use of overdamped dynamics) or the finite size of the simulation box \cite{chen}.

More generally, we believe that the wide range of relaxation scenarios predicted by our model motivates continued investigations along these lines. Given that SMCT two-correlator approaches concentrate only on temperatures above $T_c$ \cite{voigtmann_propylene,sperl}, a possible future direction would be to add an extra correlator to our model and attempt to fit data at lower temperatures. This might provide, for instance, a mechanism by which both wing and peak features may be produced in the same spectrum, as seen in the experiments of Refs \cite{kudlik} and \cite{casalini_pressure}. Although far more complicated to implement, an extension of the same approach to address non-schematic, wavevector-dependent description would also be highly desirable. 

This work was funded by EPSRC grant no.\ GR/S10377. The authors thank Th.\ Voigtmann for invaluable discussions and for supplying the time-domain algorithm; M.J.G also thanks R. A. Blythe and T. Hanney for discussions and the DTI for financial support.
M.E.C. thanks Sriram Ramaswamy for discussions.


\end{document}